\documentclass[english]{article}
\usepackage[latin9]{inputenc}
\usepackage{bbding}
\usepackage{amsmath}
\usepackage{amssymb}

\makeatletter

\DeclareFontEncoding{LGR}{}{}
\DeclareRobustCommand{\greektext}{%
  \fontencoding{LGR}\selectfont\def\encodingdefault{LGR}}
\DeclareRobustCommand{\textgreek}[1]{\leavevmode{\greektext #1}}
\ProvideTextCommand{\~}{LGR}[1]{\char126#1}

\newcommand{\lyxaddress}[1]{
	\par {\raggedright #1
	\vspace{1.4em}
	\noindent\par}
}

\makeatother

\usepackage{babel}
\begin{document}

\title{Localized states of Dirac equation }

\author{S. B. Faruque\textsuperscript{\Cross{}}, S.D. Shuvo\textsuperscript{}
and P.K. Das\textsuperscript{}}
\maketitle

\lyxaddress{\textsuperscript{}Department of Physics, Shahjalal University of
Science and Technology, Sylhet 3114, Bangladesh. }

\textsuperscript{\Cross Corresponding author. E-mail: bzaman@sust.edu}
\begin{abstract}
In this paper, we introduce an extension of the Dirac equation, very
similar to Dirac oscillator, that gives stationary localized wave
packets as eigenstates of the equation. The extension to the Dirac
equation is achieved through the replacement of the momentum operator
by a PT-symmetric generalized momentum operator. In the 1D case, the
solutions represent bound particles carrying spin having continuous
energy spectrum, where the envelope parameter defines the width of
the packet without affecting the dispersion relation of the original
Dirac equation. In the 2D case, the solutions are localized wave packets
and are eigenstates of the third component of total angular momentum
and involve Bessel functions of integral order. In the 3D case, the
solutions are localized spherical wave packets with definite total
angular momentum.
\end{abstract}
PACS number(s): 03.65.Pm

\section{Introduction}

Ordinarily, in quantum mechanics, the free particle solutions of the
Dirac equation are plane waves with infinite uncertainty in position.
But, infinite wave trains are not suitable for application. One, therefore,
creates wave packets by superposing many quantum eigenstates. Wave
packets are packets of wave function having finite width in position
and in momentum, and as such, suitable for application. The aim of
this paper is to present a particular coupling of the momentum of
a Dirac particle with a position dependent dynamical operator which
create eigenstates of the equation that are localized stationary wave
packets. This presents within the premise of Dirac theory a way alternative
to the conventional creation of wave packets by superposition of many
eigenstates- the generalized momentum operator does the job of wave
packing. This process is similar to the process of nonlinear coupling
between Coulomb motion of Rydberg electron and linearly polarized
microwave field that generate electronic wave packets as stationary
eigenstates of Schrodinger like equation {[}1-3{]}.

Study of wave packets in the context of Dirac theory is itself an
important task because of their use in nanophysics {[}4-7{]}. Moreover,
relativistic wave packet pose a challenge to theory and as such, many
authors addressed this problem{[}8-11{]}. Localized stationary wave
packets in orbit of atoms is an interesting topic being studied for
long (see, for example, {[}12{]} and the references therein). Such
stationary wave packets have numerous potential applications {[}12{]},
such as in information processing, in cavity quantum electrodynamics,
and in precision spectroscopy. The present study elevates the issue
to the relativistic regime where stationary localized wave packets
are automatic and stable products of relativistic Dirac equation.
And as such, our study opens up a new door to applications of relativistic
quantum states. Experimental realization of such stationary wave packets
can be anticipated as its predecessor Dirac oscillator has already
been realized in experiments{[}13{]}. The present work is connected
with the Dirac oscillator in the way that the Hamiltonian we employ
here is derived from Dirac oscillator{[}14-16{]}. The solutions to
Dirac oscillator are harmonic oscillator states, whereas, here we
get qualitatively very similar states, namely, wave packet states.
We present the equation and its properties in Section 2. In Section
3, we present the solutions in (1+1) freedom and discuss some of their
properties. In Section 4, we present the solution in (2+1) freedom
assuming the mass to be zero. In Section 5, we present the solution
in (3+1) freedom. Finally, in section 6, we summarize our work.

\section{Dirac equation with PT-symmetric generalized momentum}

The equation that gives wave packets as eigenstates is derived from
Dirac oscillator{[}14{]} suppressing the Dirac matrix \textgreek{b}
in the coupling operator, i.e., in the free particle Dirac equation
we replace $\overrightarrow{p}$ by $(\overrightarrow{p}-iq\overrightarrow{r})$
to obtain the following equation:
\begin{equation}
[c\overrightarrow{\alpha}.(\overrightarrow{p}-iq\overrightarrow{r})+\beta mc^{2}]\Psi=i\hbar\frac{\partial\Psi}{\partial t},
\end{equation}

where $q$ is the envelope parameter that determines the width of
the resulting wave packet, $\overrightarrow{\alpha}$, \textgreek{b}
are Dirac matrices, $\overrightarrow{p}$ and $\overrightarrow{r}$
are respectively momentum and coordinate of the fermion and c is the
speed of light in free space. This equation differs from Dirac oscillator
by only the matrix \textgreek{b} which is present in the second factor
within parentheses of the first term on the left side of Eq.(1) in
case of Dirac oscillator. The operator $\overrightarrow{(p}-iq\overrightarrow{r})$
is PT-symmetric as can be easily checked. Under P (paity) transformation:
$\overrightarrow{p}\rightarrow-\overrightarrow{p}$, $\overrightarrow{r}\rightarrow-\overrightarrow{r}$
and under T (time) reflection: $\overrightarrow{p}\rightarrow-\overrightarrow{p},\overrightarrow{r}\rightarrow\overrightarrow{r},i\rightarrow-i.$
Moreover, Eq. (1) can be generated from the free Dirac Hamiltonian
$H_{0}$ by the similarity transformation $SH_{0}S^{-1}$ with $S=exp\left(-\frac{qr^{2}}{2\hbar}\right)$.
Hence, Eq. (1) changes the description of free Dirac particles from
unlocalized states to localized states. Now, we proceed from the next
section to study its solutions. We reduce the equation in a way to
be studied in ( 1+1) freedom in the next section and find the solutions. 

\section{Localized states in one dimension}

We assume stationary solutions of Eq.(1) in the form $\Psi=\psi(\overrightarrow{r})exp(-iEt/\hbar)$.
Then, the equation becomes
\begin{equation}
[c\overrightarrow{\alpha}.(\overrightarrow{p}-iq\overrightarrow{r})+\beta mc^{2}]\psi(\overrightarrow{r})=E\psi(\overrightarrow{r}).
\end{equation}

Solutions to this equation in (1+1) freedom will be worked out with
the assumption that the motion of the particle is along the z-direction
with momentum $p$. To realize this, we use $\alpha_{z}$ in place
of $\overrightarrow{\alpha}$ and replace $\overrightarrow{r}$ by
$z$ to obtain the governing equation as,
\begin{equation}
(c\alpha_{z}p-iqc\alpha_{z}z+\beta mc^{2})\psi(z)=E\psi(z).
\end{equation}

To construct the solution of this equation, first we note that the
operator on the left of this equation, the Hamiltonian H, commutes
with the operator of z-component of spin, $\varSigma_{z}$, i.e.,
$[\varSigma_{z},H]=0.$ Hence, our solution should be simultaneous
eigenstate of energy and spin. So, we write the solution in the form
\begin{equation}
\psi(z)=\left(\begin{array}{c}
u_{1}\\
u_{2}\\
u_{3}\\
u_{4}
\end{array}\right).
\end{equation}

Inserting this into Eq.(3), we find the following coupled equations:
\begin{equation}
(cp-iqcz)u_{3}=(E-mc^{2})u_{1},
\end{equation}

\begin{equation}
(cp-iqcz)u_{1}=(E+mc^{2})u_{3},
\end{equation}

\begin{equation}
-(cp-iqcz)u_{4}=(E-mc^{2})u_{2},
\end{equation}

\begin{equation}
-(cp-iqcz)u_{2}=(E+mc^{2})u_{4}.
\end{equation}

Following traditional methods, we first assume $u_{2}=u_{4}=0$ and
find from Eqs.(5) and (6),
\begin{equation}
(cp-iqcz)(cp-iqcz)u_{1}=(E^{2}-m^{2}c^{4})u_{1},
\end{equation}

and
\begin{equation}
(cp-iqcz)(cp-iqcz)u_{3}=(E^{2}-m^{2}c^{4})u_{3}.
\end{equation}

Therefore, we see that $u_{1}$ and $u_{3}$ satisfy the same equation
and thus, will have the same structure. We can now take $u_{1}=u_{3}=0$
and find for $u_{2}$and $u_{4}$ the same governing equations as
Eq.(9) or (10). Thus, we need to solve only one equation, say, Eq.(9).
We obtain from Eq.(9),
\begin{equation}
(p^{2}-iqpz-iqzp-q^{2}z^{2})u_{1}=(\frac{E^{2}}{c^{2}}-m^{2}c^{2})u_{1}.
\end{equation}

To solve Eq.(11), we use $p=-i\hbar\frac{\partial}{\partial z}$ and
at the same time make the coordinate z dimensionless by defining a
new coordinate $z^{'}=\sqrt{\frac{q}{\hbar}}z$ and transform Eq.(11)
accordingly. In what follows only $z^{'}$ occurs and for brevity,
we drop the prime and continue to write z. Then, we find
\begin{equation}
\frac{d^{2}u_{1}}{dz^{2}}+2z\frac{du_{1}}{dz}+z^{2}u_{1}+K_{1}u_{1}=0,
\end{equation}

where
\begin{equation}
K_{1}=\frac{E^{2}}{q\hbar c^{2}}-\frac{m^{2}c^{2}}{q\hbar}+1.
\end{equation}

We now consider a solution of the form
\begin{equation}
u_{1}(z)=\phi(z)exp(-\frac{1}{2}z^{2}).
\end{equation}

Substituting this in Eq.(12), we obtain for $\phi(z)$, the governing
equation,
\begin{equation}
\frac{d^{2}\phi}{dz^{2}}+\alpha^{2}\phi=0,
\end{equation}

which immediately gives, 
\begin{equation}
\phi(z)=exp(i\alpha z),
\end{equation}

where
\begin{equation}
\alpha^{2}=\frac{E^{2}}{q\hbar c^{2}}-\frac{m^{2}c^{2}}{q\hbar}.
\end{equation}

Hence, the full solution for $u_{1}$is,
\begin{equation}
u_{1}(z)=exp(i\sqrt{\frac{q}{\hbar}}\alpha z)exp(-\frac{1}{2}\frac{q}{\hbar}z^{2}),
\end{equation}

which is a localized wave packet with the first factor giving the
oscillation in space and the second factor giving the envelope of
the packet with $q$ governing the width of the packet.The energy
associated with the wave packet is found as,
\begin{equation}
E=\pm\sqrt{\alpha^{2}c^{2}\hbar q+m^{2}c^{4}}=\pm\sqrt{\hbar^{2}k^{2}c^{2}+m^{2}c^{4}=}\pm E_{k},
\end{equation}

where $k=\sqrt{\frac{q}{\hbar}}\alpha$ is the wave number. Surprisingly,
k is independent of q as can be seen from Eq.(17). Hence, the solution
can be written more lucidly as,
\begin{equation}
u_{1}(z)=exp(ikz)exp(-\frac{1}{2}\frac{q}{\hbar}z^{2}).
\end{equation}

We now turn our attention to the spinor (4). It has the two independent
forms for spin up and spin down as follows:
\begin{equation}
\Psi_{up}(z,t)=N\left(\begin{array}{c}
1\\
0\\
1\\
0
\end{array}\right)exp(ikz)exp(-\frac{1}{2}\frac{q}{\hbar}z^{2}),
\end{equation}

\begin{equation}
\Psi_{down}(z,t)=N\left(\begin{array}{c}
0\\
1\\
0\\
1
\end{array}\right)exp(ikz)exp(-\frac{1}{2}\frac{q}{\hbar}z^{2}),
\end{equation}

The normalization factor can be calculated by demanding $\int\Psi^{\dagger}\Psi dz=1$,
which gives $N=\sqrt{\frac{1}{2}\sqrt{\frac{q}{4\pi\hbar}}}$. The
states (21) -(22) are each eigenstates of spin and energy, and represent
Gaussian wave packets with minimum uncertainty product of position
and momentum. The width of the packets are governed by the envelope
parameter q. These states have continuous energy spectrum but they
are not representing freely moving particles, rather the particles
are bound. As such, it is better to say that the particles are quasiparticles..

\section{Solution in two dimension for massless states}

To solve Eq.(2) in 2D, we use $m=0$, for that gives the theory an
opportunity to be applied to systems like graphene. And for that matter,
we use in Eq.(2) $\overrightarrow{\sigma}$ matrices in place of $\overrightarrow{\alpha}$
matrices and assume $\overrightarrow{\sigma}=(\sigma_{x},\sigma_{y}),\overrightarrow{p}=(p_{x},p_{y})$
and $\overrightarrow{r}=(x,y)=(rcos\varphi,rsin\varphi).$ The master
equation can then be written as
\begin{equation}
\left(\begin{array}{cc}
0 & cP_{-}\\
cP_{+} & 0
\end{array}\right)\left(\begin{array}{c}
\psi_{1}\\
\psi_{2}
\end{array}\right)=E\left(\begin{array}{c}
\psi_{1}\\
\psi_{2}
\end{array}\right).
\end{equation}

where
\begin{equation}
P_{+}=(p_{x}-iqx)+i(p_{y}-iqy),
\end{equation}

\begin{equation}
P_{-}=(p_{x}-iqx)-i(p_{y}-iqy).
\end{equation}

We transform Eq.(23) using polar coordinates as defined above and
use the ansatz that the solutions are eigenstates of $J_{z}=L_{z}+\frac{\hbar}{2}\sigma_{z}$
with the eigenvalues of $L_{z}$ being $m\hbar$. Hence, we write
the solution as
\begin{equation}
\Psi(r,\varphi)=\left(\begin{array}{c}
\psi_{1}\\
\psi_{2}
\end{array}\right)=e^{im\varphi}\left(\begin{array}{c}
f(r)\\
e^{i\varphi}g(r)
\end{array}\right).
\end{equation}

Using standard procedure, we obtain the second order differential
equation satisfied by $f(r)$ given by,
\begin{equation}
\frac{d^{2}f}{dr^{2}}+(\frac{1}{r}+\frac{2q}{\hbar}r)\frac{df}{dr}+\frac{q^{2}}{\hbar^{2}}r^{2}f-\frac{m^{2}}{r^{2}}f+K_{2}f=0,
\end{equation}

where $K_{2}=\frac{E^{2}}{c^{2}\hbar^{2}}+\frac{2q}{\hbar}$. Now,
an exactly similar equation is satisfied by $g(r)$ with only $m\rightarrow m+1$.
So, solving Eq.(27) only suffices for both the functions. Next, using
an alternative coordinate as defined by $r^{'}=\sqrt{\frac{q}{\hbar}}r$
and continuing with the use of $r$ for $r^{'}$, we get from Eq.(27),
\begin{equation}
\frac{d^{2}f}{dr^{2}}+(\frac{1}{r}+2r)\frac{df}{dr}+(r^{2}-\frac{m^{2}}{r^{2}}+K_{3})f=0,
\end{equation}

where $K_{3}=\frac{E^{2}}{\hbar qc^{2}}+2.$ Now, we write $f(r)=v(r)e^{-\frac{1}{2}r^{2}}$and
obtain for $v(r)$, the following equation:
\begin{equation}
\frac{d^{2}v}{dr^{2}}+\frac{1}{r}\frac{dv}{dr}+(\rho^{2}-\frac{m^{2}}{r^{2}})v=0,
\end{equation}

where $\rho^{2}=\frac{E^{2}}{\hbar qc^{2}}.$ Solution of this equation
are Bessel functions of integral order and one may choose any one
from three types of Bessel functions of integral order. Here we choose
the first kind of Bessel functions $J_{m}$ and find the solution
of Eq. (29) as $J_{m}(\rho\sqrt{\frac{q}{\hbar}r)}$. A similar calculation
yields for the functions $g(r)$ the corresponding Bessel functions
$J_{m+1}(\rho\sqrt{\frac{q}{\hbar}r)}$. Hence, we obtain the full
solution as
\begin{equation}
\Psi(r,\varphi)=Ne^{im\varphi}e^{-\frac{1}{2}\frac{q}{\hbar}r^{2}}\left(\begin{array}{c}
J_{m}(kr)\\
e^{i\varphi}J_{m+1}(kr)
\end{array}\right),
\end{equation}

where $k=\rho\sqrt{\frac{q}{\hbar}}=\frac{E}{c\hbar}$ are the wave
numbers. The dispersion relation is thus,$E=\pm\hbar kc$. Normalizatin
constant N in Eq.(30) can be evaluated using the results of Ref.{[}17{]}.
We obtain
\begin{equation}
N^{2}\frac{\pi\hbar}{q}exp\left(-\frac{\hbar k^{2}}{2q}\right)\left[I_{m}\left(\frac{\hbar k^{2}}{2q}\right)+I_{m+1}\left(\frac{\hbar k^{2}}{2q}\right)\right]=1,
\end{equation}

where $I_{m}(z)$ and $I_{m+1}(z)$ are modified Bessel functions.
The functions $\Psi(r,\varphi)$ of Eq.(30) are eigenstates of $J_{z}$,
here the total angular momentum, and of energy. The solutions in the
present case are wave packets as is evident from the structure of
Eq.(30). Moreover, the solutions (30) are similar in form as those
found for graphene quantum dots in {[}18{]} except the appearance
of the Gaussian factor in our case. Hence, we can assume that our
extension of the Dirac equation affects only the extent of the wave
function in space without affecting the energy level spectrum.

\section*{5 Solution in three dimension}

We now solve Eq.(2) in full. To do so, we decompose $\Psi(\overrightarrow{r})$as
\begin{equation}
\Psi(\overrightarrow{r})=\left(\begin{array}{c}
\psi_{1}(\overrightarrow{r})\\
\psi_{2}(\overrightarrow{r})
\end{array}\right).
\end{equation}

Using the standard representation of $\overrightarrow{\alpha}$ through
the Pauli matrices $\overrightarrow{\sigma}$ and using Eq.(32) in
Eq.(2), we obtain the two coupled equations given by
\begin{equation}
c\overrightarrow{\sigma}.\overrightarrow{p}\psi_{2}-iqc\overrightarrow{\sigma}.\overrightarrow{r}\psi_{2}=(E-mc^{2})\psi_{1},
\end{equation}

\begin{equation}
c\overrightarrow{\sigma}.\overrightarrow{p}\psi_{1}-iqc\overrightarrow{\sigma}.\overrightarrow{r}\psi_{1}=(E-mc^{2})\psi_{2}.
\end{equation}

Based on the symmetries of the Dirac equation, we use the spin-angle
functions{[}19{]} defined in two-component form as
\begin{equation}
y_{j-\frac{1}{2}}^{jm}(\hat{r})=\left(\begin{array}{c}
\sqrt{\frac{j+m}{2j}}Y_{j-\frac{1}{2}m-\frac{1}{2}}\\
\sqrt{\frac{j-m}{2j}}Y_{j-\frac{1}{2}m+\frac{1}{2}}
\end{array}\right),
\end{equation}

\begin{equation}
y_{j+\frac{1}{2}}^{jm}(\hat{r})=\left(\begin{array}{c}
-\sqrt{\frac{j-m+1}{2j+2}}Y_{j+\frac{1}{2}m-\frac{1}{2}}\\
\sqrt{\frac{j+m+1}{2j+2}}Y_{j+\frac{1}{2}m+\frac{1}{2}}
\end{array}\right),
\end{equation}

where Y's are spherical harmonics with j the total angular momentum
quantum number and m being the magnetic quantum number associated
with j. The functions given by Eqs.(35) and (36) are simultaneous
eigenfunctions of $L^{2},S^{2},J^{2},J_{z}$. Then, using standard
procedure {[}19{]}, we write
\begin{equation}
\psi_{1}(\overrightarrow{r})=u(r)y_{j-\frac{1}{2}}^{jm},
\end{equation}

and
\begin{equation}
\psi_{2}(\overrightarrow{r})=-iv(r)y_{j+\frac{1}{2}}^{jm},
\end{equation}

where in Eq.(38), the factor -i is included for later convenience.
Now, we can write {[}19{]}
\begin{equation}
\overrightarrow{\sigma}.\overrightarrow{p}=(\overrightarrow{\sigma}.\hat{r})\left[-i\hbar\frac{\partial}{\partial r}+\frac{1}{r}i\overrightarrow{\sigma}.\overrightarrow{L}\right],
\end{equation}

where $\overrightarrow{L}$ is the orbital angular momentum operator.
Now, 
\begin{equation}
(\overrightarrow{\sigma}.\overrightarrow{L})y_{l}^{jm}=\kappa y_{l}^{jm},
\end{equation}

where $\kappa=-(\lambda+1)$ for $l=j+\frac{1}{2}$ and $\kappa=(\lambda-1)$
for $l=j-\frac{1}{2}$ , where $\lambda=j+\frac{1}{2}$. It is to
be noted that {[}19{]}
\begin{equation}
\overrightarrow{\sigma}.\hat{r}y_{l=j\pm\frac{1}{2}}^{jm}=-y_{l=j\mp\frac{1}{2}}^{jm}.
\end{equation}

Inserting Eqs.(37)-(41) in Eq.(33) and rearranging, we obtain
\begin{equation}
\left(\frac{d}{dr}+\frac{\lambda+1}{r}+\frac{q}{\hbar}r\right)v(r)=\left(\frac{E-mc^{2}}{\hbar c}\right)u(r).
\end{equation}

Similarly, we obtain from Eq.(34),
\begin{equation}
\left(\frac{d}{dr}-\frac{\lambda-1}{r}+\frac{q}{\hbar}r\right)u(r)=-\left(\frac{E+mc^{2}}{\hbar c}\right)v(r).
\end{equation}

We can reduce Eqs.(42) and (43) into uncoupled form by using simple
algebra. We do this and use the dimensionless variable $r^{'}=\sqrt{\frac{q}{\hbar}}r$
and find the following equations (where we keep on using r which is
actually $r^{'}$):
\begin{equation}
\frac{d^{2}u}{dr^{2}}+\left(\frac{2}{r}+2r\right)\frac{du}{dr}+\left(r^{2}-\frac{\lambda(\lambda-1)}{r^{2}}\right)u+K_{4}u=0,
\end{equation}

\begin{equation}
\frac{d^{2}v}{dr^{2}}+\left(\frac{2}{r}+2r\right)\frac{dv}{dr}+\left(r^{2}-\frac{\lambda(\lambda+1)}{r^{2}}\right)v+K_{4}v=0,
\end{equation}

where$K_{4}=\frac{E^{2}-m^{2}c^{4}}{\hbar qc^{2}}+3.$ Using the ansatz
\begin{equation}
u(r)=\xi(r)exp\left(-\frac{1}{2}r^{2}\right),
\end{equation}

and using this in Eq.(44), we obtain
\begin{equation}
\frac{d^{2}\xi}{dr^{2}}+\frac{2}{r}\frac{d\xi}{dr}+\left(\gamma^{2}-\frac{\lambda(\lambda-1)}{r^{2}}\right)\xi=0.
\end{equation}

Similarly, using
\begin{equation}
v(r)=\chi(r)exp\left(-\frac{1}{2}r^{2}\right),
\end{equation}

we obtain from Eq.(45),
\begin{equation}
\frac{d^{2}\chi}{dr^{2}}+\frac{2}{r}\frac{d\chi}{dr}+\left(\gamma^{2}-\frac{\lambda(\lambda+1)}{r^{2}}\right)\chi=0,
\end{equation}

where in Eqs.(47) and (49), $\gamma^{2}=\frac{E^{2}-m^{2}c^{4}}{\hbar qc^{2}}.$
Solutions of Eq.(47) are spherical Bessel functions $j_{\lambda^{'}}(\gamma r)$
and $n_{\lambda^{'}}(\gamma r)$ , where, $\lambda^{'}=\lambda-1.$
Solutions of Eq.(49) are also spherical Bessel functions, namely,
$j_{\lambda}(\gamma r)$ and $n_{\lambda}(\gamma r)$. Hence, we can
write explicitly, using only regular solutions and restoring the original
variable r,
\begin{equation}
\psi_{1}(\overrightarrow{r})=Nj_{\lambda^{'}}(kr)exp\left(-\frac{1}{2}\frac{q}{\hbar}r^{2}\right)y_{\lambda^{'}}^{jm},
\end{equation}

and
\begin{equation}
\psi_{2}(\overrightarrow{r})=-iNj_{\lambda}(kr)exp\left(-\frac{1}{2}\frac{q}{\hbar}r^{2}\right)y_{\lambda}^{jm},
\end{equation}

where
\begin{equation}
E^{2}=\hbar^{2}k^{2}c^{2}+m^{2}c^{4},
\end{equation}

or,
\begin{equation}
k=\pm\frac{\sqrt{E^{2}-m^{2}c^{4}}}{\hbar c}.
\end{equation}

Evidently, k is independent of q, the envelope parameter. Next, we
normalize the wavefunction (32) with $\psi_{1}$and $\psi_{2}$ given
by Eqs.(50) and (51) where N is the normalization constant. Using
results of Ref.{[}17{]} , we obtain
\begin{equation}
N^{2}\left(\frac{\hbar}{q}\right)^{\frac{3}{2}}\frac{\sqrt{\pi}}{4}exp\left(-\frac{\hbar k^{2}}{2q}\right)\left[f_{\lambda^{'}}\left(\frac{\hbar k^{2}}{2q}\right)+f_{\lambda}\left(\frac{\hbar k^{2}}{2q}\right)\right]=1,
\end{equation}

where $f_{n}(z)=\sqrt{\frac{\pi}{2z}}I_{n+\frac{1}{2}}(z)$ are the
modified spherical Bessel function of the first kind. For completeness,
we now write explicitly the spinor for $j=\frac{3}{2}$ and $m=\frac{3}{2},$which
is
\begin{equation}
\Psi^{\frac{3}{2}\frac{3}{2}}=Nexp\left(-\frac{1}{2}\frac{q}{\hbar}r^{2}\right)\left(\begin{array}{c}
j_{1}(kr)Y_{11}\\
0\\
ij_{2}(kr)\sqrt{\frac{1}{5}}Y_{21}\\
-ij_{2}(kr)\sqrt{\frac{4}{5}}Y_{22}
\end{array}\right).
\end{equation}

The solutions found, namely, Eq.(32) with $\psi_{1}$and $\psi_{2}$
given by Eqs.(50) and (51) with the specific example given by Eq.(55)
are wave packets in three dimensions carrying total angular momentum
and its z-component given by the quantum numbers $j$ and $m$ as
conserved quantities. Hence, we get here stationary spherical Bessel
wave packets carrying angular momentum. Thus, we have found a complete
picture of the solutions of Eq.(2) which is an extension of the Dirac
equation very similar to Dirac oscillator.

\section*{6 Summary and conclusion}

In this paper, we have presented an extension of the Dirac equation,
very similar to the Dirac oscillator, given by Eq.(1). We have solved
the (1+1) case of the equation and the solutions are given by Eqs.(21)-(22).
The solutions are spinor wave packets carrying definite spin (1/2
or -1/2) and continuous energy spectrum. They represent bound quasiparticles
although the spectrum is continuous. The states are of minimum position-momentum
uncertainty product, the widths in position and momentum space are
determined by the envelope parameter q. This parameter entering Eq.(1)
via the operator $(-iq\overrightarrow{r})$ does not affect the dispersion
relation, given by Eq.(19), but only packs the otherwise sinusoidal
waves into a Gaussian envelope. This is why we call Eq.(1) the ``wave
packing Dirac equation''. The solutions can also be looked at as
representing freely moving particles, but in that case they will suffer
dispersion owing to the nonlinear dispersion relation (19). Extension
of the system to the massless case is easy and the solutions remain
same. Finally, in section 5, we have solved the equation in full,
using spherical polar coordinates and spin-angle functions. The same
envelope function shows up and it envelopes the free particle functions
given by spherical Bessel functions. In all the three cases, namely,
one to three dimensions, we obtain wave packets carrying angular momentum.
In the 1D case, it is the spin that is conserved; in the 2D case,
it is the total angular momentum, which in this case is the z-component
of total angular momentum, is conserved. In the 3D case, it is the
total angular momentum and its z-component that are conserved quantities
in the solution. And in all the cases, the solutions are stationary
wave packets. In conclusion, we have found a procedure to localize
the otherwise unlocalized free Dirac spinors by transforming the free
Dirac Hamiltonian to a new form via a similarity transformation.

\section*{Acknowledgment}

This work is supported by Grant No. PS/2018/2/22 endowed by the Research
Centre of Shahjalal University of Science and Technology, Sylhet,
Bangladesh.

\end{document}